\documentclass{ws-p8-50x6-00}

\begin{document}
%\draft 

\title{On the accuracy of the post-Newtonian approximation\footnote{In
``2001: a relativistic spacetime odyssey'', Proc. of the 25th Johns
Hopkins workshop, edited by Ignacio Ciufolini, Daniele Dominici and
Luca Lusanna, World Scientific, p. 411 (2001).}}

\author{Luc Blanchet} 

\address{Gravitation et Cosmologie (GReCO),\\ Institut d'Astrophysique
de Paris -- C.N.R.S.,\\ 98\textsuperscript{~$\!bis$} boulevard Arago,
75014 Paris, France\\
E-mail: blanchet@iap.fr} 

%\date{\today}
%\twocolumn[

\maketitle
%\widetext

\abstracts{We apply standard post-Newtonian methods in general
relativity to locate the innermost circular orbit (ICO) of
irrotational and corotational binary black-hole systems. We find that
the post-Newtonian series converges well when the two masses are
comparable. We argue that the result for the ICO which is predicted by
the third post-Newtonian (3PN) approximation is likely to be very
close to the ``exact'' solution, within 1\% of fractional accuracy or
better. The 3PN result is also in remarkable agreement with a
numerical calculation of the ICO in the case of two corotating black
holes moving on exactly circular orbits. The behaviour of the
post-Newtonian series suggests that the gravitational dynamics of two
bodies of comparable masses does not resemble that of a test particle
on a Schwarzschild background. This leads us to question the validity
of some post-Newtonian resummation techniques that are based on the
idea that the field generated by two black holes is a deformation of
the Schwarzschild space-time.}

%\pacs{04.30.-w, 04.80.Nn, 97.60.Jd, 97.60.Lf}
%]

%\narrowtext

\section{Introduction}

The ``standard'' post-Newtonian approximation, or expansion when the
speed of light $c\to +\infty$, is at the basis of an important body of
research, which has provided us in the past with our best picture of
the motion of compact objects in general relativity. We quote the
pionneering works of Einstein\cite{einstein}, Droste\cite{droste}, and
DeSitter\cite{desitter}, the landmark analysis due to Einstein, Infeld
and Hoffmann\cite{EIH} of the dynamics of $N$ separated bodies at the
first post-Newtonian order (1PN, or $1/c^2$), and the seminal papers
by Chandrasekhar and collaborators\cite{C65,CN69,CE70} concerning the
equations of motion of extended fluid systems, up to the 2.5PN level
(at which order appears the first effect due to the reaction to the
emission of gravitational radiation). In the case of two compact
objects (neutron stars or black holes), we possess the 2.5PN equations
of motion of the binary pulsar\cite{DD81a,D82,Dhouches,BFP98}, and the
3PN equations of motion of inspiralling compact
binaries\cite{JaraS98,JaraS99,DJS00,DJS01,DJSdim,BF00,BFreg,BFregM,BFeom,ABF01}.
Regarding the problem of the gravitational radiation emitted by
inspiralling compact binaries, we have under control most of the
gravitational-wave flux up to the 3.5PN
order\cite{BDI95,WWi96,BIJ02,BFIJ02}, including the specific effects
of wave tails\cite{B96,B98tail}.

In this paper we focus our attention on the question of the dynamics
of black-hole binary systems (henceforth we assume that the compact
objects are black holes) and its recent resolution up to the 3PN
approximation, corresponding to the formal order $1/c^6$ beyond the
Newtonian force law. On the one hand the ADM-Hamiltonian formalism of
general relativity has been applied at the 3PN order by Jaranowski and
Sch\"afer\cite{JaraS98,JaraS99}, and Damour, Jaranowski and
Sch\"afer\cite{DJS00,DJS01}. On the other hand a direct 3PN iteration
of the equations of motion --- instead of a Hamiltonian --- in
harmonic coordinates (extending the method proposed in
Ref.\cite{BFP98}) has been implemented by Blanchet and
Faye\cite{BF00,BFreg,BFregM,BFeom}, and de Andrade, Blanchet and
Faye\cite{ABF01}. These two independent approaches have succeeded; it
has been shown that there exists a unique transformation of the
particle's dynamical variables that changes the 3PN
harmonic-coordinates Lagrangian\cite{ABF01} into a Lagrangian whose
Legendre transform is exactly identical to the 3PN ADM-coordinates
Hamiltonian\cite{DJS00}.

In the previous approaches the two black holes are modelled by point
particles, solely described by their masses $m_1$ and $m_2$. This is
consistent with the very spirit of the post-Newtonian method. The
point-particle description has to be supplemented by a process of
regularization of the self field of point particles. The standard
regularization {\it \`a la} Hadamard is at the basis of the
works\cite{JaraS98,JaraS99,DJS00,DJS01,BF00,BFeom,ABF01} (an extended
version of this regularization has been defined in
Refs.\cite{BFreg,BFregM}).  Unfortunately it was shown that at the 3PN
order the equations of motion of black holes contain a particular
coefficient, i.e. the ``static'' ambiguity $\omega_{\rm s}$ in the
ADM-Hamiltonian formalism\cite{JaraS98,JaraS99,DJS00,DJS01} or the
parameter $\lambda$ in the harmonic-coordinates
approac\cite{BF00,BFreg,BFregM,BFeom,ABF01}\footnote{See Eq. (\ref{5})
below for the relation linking together $\lambda$ and $\omega_{\rm
s}$.}, which cannot be fixed by the Hadamard regularization. The value
of this coefficient has been obtained by means of a dimensional
regularization instead of the Hadamard one within the ADM-Hamiltonian
formalism\cite{DJSdim}. We shall discuss some implications of this
result (i.e. $\omega_{\rm s}=0$) regarding the validity of the
post-Newtonian expansion.

Let us look at the conserved energy of the black-hole binary in the
center-of-mass frame at the 3PN order. Technically this energy follows
from the 3PN harmonic-coordinates Lagrangian\cite{ABF01} or
equivalently from the 3PN ADM-coordinates Hamiltonian\cite{DJS00}. The
energy is conserved only when we neglect the radiation reaction
damping at the 2.5PN order. The time derivative of the energy is equal
to the radiation reaction effect, but for the present discussion we
are interested only in the conservative part of the dynamics which is
composed of the Newtonian, 1PN, 2PN and 3PN
approximations. Specializing to the case of orbits which are circular
(apart from the gradual radiation-reaction inspiral)\footnote{We know
that most inspiralling compact binaries have circular orbits because
of radiation-reaction effects.}  yields then the center-of-mass energy
$E$ at the 3PN order for circular orbits as a function of the
frequency $\omega$ of the orbital motion.

Here we want to assess the validity of the post-Newtonian
approximation. More precisely we address, and to some extent we
answer, the following questions. How accurate is the post-Newtonian
expansion for describing the dynamics of binary black hole systems~?
Is the innermost circular orbit (ICO) of binary black holes, defined
by the minimum of the energy function $E(\omega)$, accurately
determined at the highest currently known post-Newtonian order~? This
question is pertinent because the ICO represents a point in the late
stage of evolution of the binary which is very relativistic (orbital
velocities of the order of 50\% of the speed of light). How well does
the 3PN approximation as compared with the prediction provided by
numerical relativity~? What is the validity of the various
post-Newtonian resummation techniques\cite{DIS98,BD01} which aim at
``boosting'' the convergence of the standard post-Newtonian
approximation~?

The previous questions are very interesting but difficult to settle
down rigorously. Indeed the very essence of an approximation method is
to cope with our ignorance of the higher-order terms in some
expansion, but the higher-order terms are precisely the ones which
would be needed for a satisfying answer to these problems. So we shall
be able to give only some educated guesses and/or plausible answers,
that we cannot justify rigorously, but which seem very likely from the
standard point of view on the post-Newtonian theory, in particular
that the successive orders of approximation get smaller and smaller as
they should (in average), with only few accidents occuring at high
orders where a particular approximation would be abnormally large with
respect to the lower-order ones. Admittedly, in addition, our faith in
the estimation we shall give regarding the accuracy of the 3PN order
for instance, comes from the historical perspective, thanks to the
many successes achieved in the past by the post-Newtonian
approximation when confronting the theory and observations (e.g. of
the Solar system dynamics and the binary pulsar). It is indeed beyond
question, from our past experience, that the post-Newtonian method
does work.

There are many other related issues that we shall not address. For
instance~: is the best presently known post-Newtonian approximation,
i.e. 3PN, sufficient for the problem of the coalescence of two black
holes in a foreseeable future\footnote{We have in mind the 3PN
approximation considered as an input for the numerical calculation of
the black-hole coalescence (assuming that a clear method would exist
for implementing the post-Newtonian initial conditions into a
numerical scheme).}~? A related problem is the qualitative and
quantitative influence of the high-order radiation reaction effects
which have to be superposed to the conservative part of the
dynamics. Finally we shall discuss only the binary's equations of
motion, and leave aside the problem of the radiation field. Certainly
the accuracy of the 3.5PN gravitational-radiation flux of black-hole
binaries which has been computed in Refs.\cite{BIJ02,BFIJ02} should
be discussed in a similar way.

Basically, the point we would like to emphasize\footnote{We are
following the recent study in Ref.\cite{B02ico}.} is that the
post-Newtonian approximation, in standard form (without using the
resummation techniques advocated in Refs.\cite{DIS98,BD01}), is able
to located the ICO of two black holes with an excellent accuracy of
the order of 1\%, and perhaps much better than that. At first sight
this statement is rather surprising, because the dynamics of two black
holes at the point of the ICO is so relativistic. Indeed one often
ears about the ``bad convergence'', or the ``fundamental breakdown'',
of the post-Newtonian series in the regime of the ICO. However our
estimates do show that the 3PN approximation is very good in this
regime, and they are also confirmed by the remarkable agreement, we
shall detail below, with a numerical calculation by Gourgoulhon {\it
et al.}\cite{GGB1,GGB2} of the ICO in the case of two black holes
moving on exactly circular orbits (``helical symmetry''). When
comparing the post-Newtonian approximation with the numerical
simulation we face an interesting problem~: since the numerical work
has been done for corotating black holes, which spin with the orbital
velocity $\omega$, the effects of spins, appropriate to two Kerr black
holes, are to be taken into account within our post-Newtonian
framework.

Another point we shall develop is that most probably the
general-relativistic dynamics of two objects with comparable masses is
qualitatively different (in addition of being far more complicated),
than the dynamics of a ``test'' particle with geodesic motion on a
fixed background Schwarzschild space-time. Our argument has something
to do with the value $\omega_{\rm s}=0$ obtained in Ref.\cite{DJSdim}
for the 3PN regularization ambiguity parameter. Indeed this value, if
we believe that it is really the prediction of general relativity,
suggests that the two-body interaction is {\it not}
``Schwarzschild-like'', in the sense that it does not seem to admit a
light-ring singularity similar to the one of the Schwarzschild
space-time. We then argue that this fact sheds a doubt on the
validity, at high post-Newtonian orders, of some resummation
techniques (like Pad\'e approximants\cite{DIS98}) that are based on
the assumption that the field generated by two bodies of comparable
sizes is a ``deformation'' of the Schwarzschild metric. {\it A
contrario} we shall find that the value $\omega_{\rm s}=0$ gives some
convincing evidence that the standard post-Newtonian approximation,
based on Taylor rather than Pad\'e approximants, is very accurate.

\section{The binding energy of two black-holes}

The binding energy, in the center-of-mass frame, is defined as the
invariant energy associated with the {\it conservative} part of the
binary's 3PN dynamics (we ignore the radiation reaction effect at the
2.5PN order). Restricting our consideration to circular orbits, the
energy is a function of a single variable, which can be chosen to be
the distance $r$ between the two particles in a given coordinate
system, or, better, the frequency $\omega=\frac{2\pi}{P}$ of the
orbital motion ($P$ is the orbital period). We introduce for
convenience the particular frequency-related parameter

\begin{equation}\label{1}
x \equiv \left(\frac{G M \omega}{c^3}\right)^{2/3}\;.
\end{equation}
The mass-energies of the black holes are $m_1$ and $m_2$ (they take
notably into account the rotational energies); the total mass is
$M=m_1+m_2$. It is important to express the energy in terms of the
frequency-related parameter $x$, instead of some coordinate distance
$r$, because the function $E(x)$ then takes an invariant expression
(the same in different coordinate systems).

Having in hands the circular-orbit energy, we then define the
innermost circular orbit (ICO) as the minimum, when it exists, of the
energy function $E(x)$. The definition is motivated by our comparison
with the numerical calculation\cite{GGB1,GGB2}, because this is
precisely that minimum which is computed numerically. In particular,
we do not define the ICO as a point of dynamical general-relativistic
unstability. The circular-orbit energy, developed to the 3PN order, is
of the form

\begin{equation}\label{2}
E(x) = M c^2 -\frac{\mu\,c^2 x}{2} \Big\{ 1 + a_1(\nu)\,x +
a_2(\nu)\,x^2 + a_3(\nu)\,x^3+{\cal O}(x^4)\Big\}\;.
\end{equation} 
The first term is the rest-mass, the next one, proportional to $x$, is
the Newtonian term, and then we have many post-Newtonian corrections,
the coefficients of which are known at present only up to the 3PN
order\cite{JaraS98,JaraS99,DJS00,DJS01,BF00,BFreg,BFregM,BFeom,ABF01},
and given by

\begin{eqnarray}
a_1(\nu) &=& -\frac{3}{4}-\frac{\nu}{12}\;,\label{3a}\\ a_2(\nu) &=&
-\frac{27}{8}+\frac{19}{8}\nu -\frac{\nu^2}{24}\;,\label{3b}\\ a_3(\nu) &=&
-\frac{675}{64}+\left[\frac{209323}{4032}-\frac{205}{96}\pi^2
-\frac{110}{9}\lambda\right]\nu-\frac{155}{96}\nu^2
-\frac{35}{5184}\nu^3\;.\label{3c}
\end{eqnarray} 
We make use of the useful ratio between the reduced and total masses~:

\begin{equation}\label{4}
\nu = \frac{\mu}{M} \quad\hbox{where}\quad \mu = \frac{m_1 m_2}{M}\;.
\end{equation}
This ratio is interesting because of its range of
variation,

\begin{equation}\label{4'}
0<\nu\leq \frac{1}{4}\;,
\end{equation}
where $\nu=\frac{1}{4}$ in the equal-mass case and $\nu\to 0$ in the
test-mass limit for one of the bodies. We shall investigate the value
of the ICO as predicted by the 3PN energy (\ref{2})-(\ref{3c}) in
Section 4.

The 3PN coefficient $a_3(\nu)$ involves the regularization-ambiguity
parameter $\lambda$ introduced in Refs.\cite{BF00,BFeom} and due to a
physical incompleteness in the Hadamard method\cite{BFreg,BFregM} for
regularizing the self-field of point particles. This parameter is
equivalent to the parameter $\omega_{\rm s}$ in
Refs.\cite{JaraS98,JaraS99} and related to it
by\cite{BF00,DJS01,ABF01}~:

\begin{equation}\label{5}
\lambda = -\frac{3}{11} \omega_{\rm s}-\frac{1987}{3080}\;.
\end{equation}
It has been argued in Ref.\cite{DJSisco} that the numerical value of
$\omega_{\rm s}$ could be $\simeq -9$, because for such a value some
different resummation techniques, when they are implemented at the 3PN
order, give approximately the same result for the ICO. Even more, it
was suggested\cite{DJSisco} that $\omega_{\rm s}$ might be precisely
equal to $\omega_{\rm s}^*$, with

\begin{equation}\label{6}
\omega_{\rm s}^*=-\frac{47}{3}+\frac{41}{64}\pi^2 = -9.34\cdots\;.
\end{equation}
However, a more recent computation performed with the help of a
dimensional regularization instead of the Hadamard regularization,
within the ADM-Hamiltonian formalism\cite{DJSdim}, has yielded

\begin{equation}\label{7}
\omega_{\rm s}=0~~\Longleftrightarrow~~\lambda=-\frac{1987}{3080}\;.
\end{equation}
Here we adopt $\omega_{\rm s}=0$ as the ``correct'' value predicted by
general relativity for the ambiguity parameter. Note that the result
(\ref{7}) is quite different from $\omega_{\rm s}^*$ given by
Eq. (\ref{6})~: this suggests, according to Ref.\cite{DJSisco}, that
different resummation techniques, {\it viz} Pad\'e
approximants\cite{DIS98} and effective-one-body methods\cite{BD01},
which are designed to ``accelerate'' the convergence of the
post-Newtonian series, do not in fact converge toward the same exact
solution (or, at least, not as fast as expected).

The appearance of one ambiguity parameter at the 3PN order is
interesting in connection with the so-called effacing principle
satisfied by general relativity\cite{Dhouches}, according to which
the equations of motion and the radiation field of gravitationally
condensed objects should depend only on their relativistic masses and
not on the detailed features of their internal structure --- we
neglect the tidal effects between the objects. In general relativity
this principle follows from the strong equivalence principle (which
differs from the Einstein equivalence principle by the inclusion of
bodies with self-gravitational interactions and of experiments
involving gravitational forces). Indeed, in the freely falling frame
of one of the (spherically symmetric) compact objects, we can apply
the Birkhoff theorem and find that the external vacuum field depends
only on the mass. In consequence we should expect that the parameter
$\lambda$ is a pure number (e.g. a rational fraction), independent of
the internal structure of the compact objects.

It would be interesting to confirm Eq. (\ref{7}) by an independent
calculation, hopefully without resort to any regularization scheme. We
have in mind a calculation valid for extended (``fluid'') systems,
taking {\it a priori} into account the internal structure of the
fluids. Such a calculation would provide, in addition to the
determination of $\lambda$, an explicit verification of the effacing
principle of general relativity. Notice that from the point of view of
a calculation valid for extended fluids, it is hard to believe that
the ambiguity parameter could depend on $\pi^2$ like in Eq. (\ref{6}).

In the limiting case $\nu\to 0$, the energy (\ref{2})-(\ref{3c})
reduces to the 3PN approximation of the exact expression for a test
particle in the Schwarzschild background,

\begin{equation}\label{8}
E^{\rm Sch}(x) = \mu\,c^2\,\frac{1-2x}{\sqrt{1-3x}}\;.
\end{equation}
The minimum of that function or ICO occurs at $x^{\rm Sch}_{\rm
ICO}=\frac{1}{6}$, and we have $E^{\rm Sch}_{\rm ICO}=\mu
c^2\Big(\sqrt{\frac{8}{9}}-1\Big)$. We know that the ICO in the case
of the Schwarzschild metric is also an innermost {\it stable} circular
orbit (or ISCO), i.e. it corresponds to a point of dynamical
unstability. Another important feature of Eq. (\ref{8}) is the
singularity at the value $x^{\rm Sch}_{\rm light-ring}=\frac{1}{3}$
which corresponds to the famous circular orbit of photons in the
Schwarzschild metric (``light-ring'' singularity). This orbit can also
be viewed as the last {\it unstable} circular orbit. We can check that
the post-Newtonian coefficients $a_n^{\rm Sch}\equiv a_n(0)$
corresponding to Eq. (\ref{8}) are given by

\begin{equation}\label{9}
a_n^{\rm Sch} = - \frac{3^n(2n-1)!!(2n-1)}{2^n(n+1)!}\;.
\end{equation}
They increase with $n$ by roughly a factor 3 at each order. This is
simply the consequence of the fact that the radius of convergence of
the post-Newtonian series is given by the Schwarzschild light-ring
singularity at the value $\frac{1}{3}$. We may therefore recover the
light-ring orbit by investigating the limit

\begin{equation}\label{10}
\lim_{n\to +\infty}\,\frac{a_{n-1}^{\rm Sch}}{a_n^{\rm Sch}} \,=\,
\frac{1}{3}\,=\,x^{\rm Sch}_{\rm light-ring}\;.
\end{equation}

\section{Accuracy of the post-Newtonian expansion}

Let us now discuss the accuracy of the 3PN approximation when
estimating the ICO of binary black holes. First of all we make a few
order-of-magnitude estimates. At the location of the ICO we shall find
(see the next Section) that the frequency-related parameter $x$
defined by Eq. (\ref{1}) is approximately of the order of
20\%. Therefore, we might {\it a priori} expect that the contribution
of the 1PN approximation to the energy at the point of the ICO should
be of that order. For the present discussion we take the pessimistic
view that the order of magnitude of an approximation represents also
the order of magnitude of the higher-order terms which are
neglected. We see that the 1PN approximation should yield a rather
poor estimate of the ``exact'' result, but this is quite normal at
this very relativistic point where the orbital velocity is
$\frac{v}{c}\sim x^{1/2}\sim 50\%$. By the same argument we infer that
the 2PN approximation should do much better, with fractional errors of
the order of $x^2\sim 5\%$, while 3PN will be even better, with the
accuracy $x^3\sim 1\%$.

The simple order-of-magnitude estimate suggests therefore that the 3PN
order should be close to the ``exact'' solution for the ICO to within
1\% of fractional accuracy. We think that this is very good, and we
should even remember that this estimate is pessimistic, because we can
reasonably expect that the neglected higher-order approximations, 4PN
and so on, are in fact smaller numerically (e.g. of the order of
$x^4\sim 0.2\%$). But let us keep for the present discussion the 1\%
guess for the accuracy of the 3PN approximation.

Now the previous estimate makes sense only if the numerical values of
the post-Newtonian coefficients in Eqs. (\ref{3a})-(\ref{3c}) stay
roughly of the order of one. If this is not the case, and if the
coefficients increase dangerously with the post-Newtonian order $n$,
one sees that the post-Newtonian approximation might in fact be very
bad. It has often been emphasized in the litterature (see
e.g. Refs.\cite{3mn,P95,DIS98}) that in the test-mass limit $\nu\to
0$ the post-Newtonian series converges slowly, so the post-Newtonian
approximation is not very good in the regime of the ICO. Indeed we
have seen that when $\nu=0$ the radius of convergence of the series is
$\frac{1}{3}$ (not so far from $x^{\rm Sch}_{\rm ICO}=\frac{1}{6}$),
and that accordingly the post-Newtonian coefficients increase by a
factor $\sim 3$ at each order. So it is perfectly correct that in the
case of test particles in the Schwarzschild background the
post-Newtonian approximation is to be carried out to a high order in
order to locate the turning point of the ICO.

Let us immediately remark that this negative conclusion does not
matter~: indeed we shall never use the post-Newtonian approximation
when $\nu\to 0$ simply because we know the exact result which is given
by Eq. (8)\footnote{The exact result for the radiation field is also
known, albeit only numerically.}. Therefore we should not worry about
the poor convergence of the post-Newtonian series in the test-mass
limit. The post-Newtonian method is useless and even one might say
irrelevant when considering the motion of a test particle around a
Schwarzschild black hole.

What happens when the two masses are comparable ($\nu=\frac{1}{4}$)~?
It is clear that the accuracy of the post-Newtonian approximation
depends crucially on how rapidly the post-Newtonian coefficients
increase with $n$. We have seen that in the case of the Schwarzschild
metric the latter increase is in turn related to the existence of a
light-ring orbit. For continuing the discussion we shall say that the
relativistic interaction between two bodies of comparable masses is
``Schwarzschild-like'' if the post-Newtonian coefficients
$a_n(\frac{1}{4})$ increase when $n\to +\infty$. If this is the case
this signals the existence of something like a light-ring singularity
which could be interpreted as the deformation, when the mass ratio
$\nu$ is ``turned on'', of the Schwarzschild light-ring orbit. By
analogy with Eq. (\ref{10}) we can estimate the location of this
``pseudo-light-ring'' orbit by

\begin{equation}\label{11}
\frac{a_{n-1}(\nu)}{a_n(\nu)} \sim \,x_{\rm
light-ring}(\nu)\quad\hbox{with $n=3$}\;.
\end{equation}
Here $n=3$ is the highest known post-Newtonian order. If the two-body
problem is ``Schwarzschild-like'' then the right-hand-side of
Eq. (\ref{11}) is small (say around $\frac{1}{3}$), the post-Newtonian
coefficients typically increase with $n$, and most likely it should be
difficult to get a reliable estimate by post-Newtonian methods of the
location of the ICO. So we ask~: is the gravitational interaction
between two comparable masses Schwarzschild-like~?

\begin{table}
\caption{Numerical values of the sequence of coefficients of the
post-Newtonian series composing the energy function
(\ref{2})-(\ref{3c}).\label{tab1}}
\begin{center}
\footnotesize
\begin{tabular}{lccccc}
&Newtonian&$a_1(\nu)$&$a_2(\nu)$&$a_3(\nu)$\\[1mm]\hline\hline $\nu=0$
&1&-0.75&-3.37&-10.55\\[0.5mm]\hline $\nu=\frac{1}{4}$ $\quad\omega^*_{\rm
s}\simeq -9.34$&1&-0.77&-2.78&-8.75\\[0.5mm]\hline $\nu=\frac{1}{4}$
$\quad\omega_{\rm s}=0$&1&-0.77&-2.78&-0.97\\
\end{tabular}
\end{center}
\end{table}

In Table \ref{tab1} we present the values of the coefficients
$a_n(\nu)$ in the test-mass limit $\nu=0$ [see Eq. (\ref{9}) for the
analytic expression], and in the equal-mass case $\nu=\frac{1}{4}$
when the ambiguity parameter takes the ``uncorrect'' value
$\omega^*_{\rm s}$ [defined by Eq. (\ref{6})] and the correct one
$\omega_{\rm s}=0$ predicted by general relativity. When $\nu=0$ we
clearly see the expected increase of the coefficients by roughly a
factor 3 at each step. Now when $\nu=\frac{1}{4}$ and $\omega_{\rm
s}=\omega^*_{\rm s}$ we notice that the coefficients increase
approximately in the same manner as in the test-mass case
$\nu=0$. This indicates that the gravitational interaction in the case
of $\omega^*_{\rm s}$ looks like that in a one-body problem. The
associated pseudo-light-ring singularity is estimated using
Eq. (\ref{11}) as

\begin{equation}\label{11'}
x_{\rm light-ring}(\hbox{$\frac{1}{4}$},{\omega^*_{\rm s}})\sim
0.32\;.
\end{equation}
The pseudo-light-ring orbit seems to be a very small deformation of
the Schwarzschild light-ring orbit given by Eq. (\ref{10}). In this
Schwarzschild-like situation, we should not expect the post-Newtonian
series to be very accurate.

Now in the case $\nu=\frac{1}{4}$ but when the ambiguity parameter
takes the correct value $\omega_{\rm s}=0$, we see that the 3PN
coefficient $a_3(\frac{1}{4})$ is of the order of minus one instead of
being $\sim -10$. This suggests (unless 3PN happens to be quite
accidental) that the post-Newtonian coefficients in general relativity
do not increase very much with $n$. This is an interesting finding
because it indicates that the actual two-body interaction in general
relativity is {\it not} Schwarzschild-like. There does not seem to
exist something like a light-ring orbit which would be a deformation
of the Schwarzschild one. Applying Eq. (\ref{11}) we obtain as an
estimate of the ``light-ring''~:

\begin{equation}\label{11''}
x_{\rm light-ring}(\hbox{$\frac{1}{4}$},\hbox{G.R.}) \sim 2.88\;.
\end{equation}
It is clear that if we believe the correctness of this estimate we
must conclude that there is in fact {\it no} notion of a light-ring
orbit in the real two-body problem. Or, one might say (pictorially
speaking) that the light-ring orbit gets hidden inside the horizon of
the final black-hole formed by coalescence. Furthermore, if we apply
Eq. (\ref{11}) using the 2PN approximation $n=2$ instead of the 3PN
one $n=3$, we get the value $\sim 0.28$ instead of
Eq. (\ref{11''}). So at the 2PN order the metric seems to admit a
light ring, while at the 3PN order it apparently does not admit
any. This erratic behaviour reinforces our idea that it is meaningless
(with our present 3PN-based knowledge, and untill fuller information
is available) to assume the existence of a light-ring singularity when
the masses are equal.

It is impossible of course to be thoroughly confident about the
validity of the previous statement because we know only the
coefficients up to 3PN order. Any tentative conclusion based on 3PN
can be ``falsified'' when we obtain the next 4PN order. Nevertheless,
we feel that the mere fact that $a_3(\frac{1}{4})=-0.97$ in Table
\ref{tab1} is sufficient to motivate our (tentative) conclusion that
the gravitational field generated by two bodies is more complicated
than the Schwarzschild space-time. This appraisal should look cogent
to relativists and is in accordance with the author's respectfulness
of the complexity of the Einstein field equations.

We want next to comment on a possible implication of our conclusion as
regards the so-called post-Newtonian resummation techniques,
i.e. Pad\'e approximants\cite{DIS98,DJSisco}, which aim at
``accelerating'' the convergence of the post-Newtonian series in the
pre-coalescence stage, and effective-one-body (EOB)
methods\cite{BD01,DJSisco}, which attempt at describing the late stage
of the coalescence of two black holes. These techniques are based on
the idea that the gravitational two-body interaction is a
``deformation'' --- with $\nu\leq\frac{1}{4}$ being the deformation
parameter --- of the Schwarzschild space-time. The Pad\'e approximants
are valuable tools for giving accurate representations of functions
having some singularities. In the problem at hands they would be
justified if the ``exact'' expression of the energy [whose 3PN
expansion is given by Eqs. (\ref{2})-(\ref{3c})] would admit a
singularity at some reasonable value of $x$ (e.g. $\leq 0.5$). In the
Schwarzschild case, for which Eq. (\ref{10}) holds, the Pad\'e series
converges rapidly\cite{DIS98}~: the Pad\'e constructed only from the
2PN approximation of the energy --- keeping only $a_1^{\rm Sch}$ and
$a_2^{\rm Sch}$ --- already coincide with the exact result given by
Eq. (\ref{8}). On the other hand, the EOB method maps the
post-Newtonian two-body dynamics (at the 2PN or 3PN orders) on the
geodesic motion on some effective metric which happens to be a
$\nu$-deformation of the Schwarzschild space-time. In the EOB method
the effective metric looks like Schwarzschild {\it by definition}, and
we might expect the two-body interaction to own some
Schwarzschild-like features.

Our comment is that the validity of these post-Newtonian resummation
techniques is questionable because the value $\omega_{\rm s}=0$
suggests that the two-body interaction is not Schwarzschild-like. Our
doubt is confirmed by the finding of Ref.\cite{DJSisco} (already
alluded to above) that in the case of the ``wrong'' ambiguity
parameter $\omega^*_{\rm s}\simeq -9.34$ the Pad\'e approximants and
the EOB method at the 3PN order give the same result for the ICO. From
the previous discussion we see that this agreement is to be expected
because a deformed light-ring singularity seems to exist with
$\omega^*_{\rm s}$. By contrast, in the case of general relativity
($\omega_{\rm s}=0$), the Pad\'e and EOB methods give quite different
results ({\it cf.} the figure 2 in Ref.\cite{DJSisco}). Such a
disagreement, we argue, is due to the fact that the assumptions
underlying the various resummation techniques are probably not
fulfilled, so they may converge toward different solutions. Another
confirmation comes from the light-ring singularity which is determined
from the Pad\'e approximants at the 2PN order [see Eq. (3.22) in
Ref.\cite{DIS98}] as

\begin{equation}\label{11'''}
x_{\rm light-ring}(\hbox{$\frac{1}{4}$},\hbox{Pad\'e})\sim 0.44\;.
\end{equation}
This value is rather close to Eq. (\ref{11'}) but strongly disagrees
with Eq. (\ref{11''}). Our explanation is that the Pad\'e series
converges toward a theory having $\omega_{\rm s}\simeq\omega^*_{\rm
s}$ and so which is different from general relativity.

Finally we come to the good news that, if really the (absolute value
of the) post-Newtonian coefficients when $\nu=\frac{1}{4}$ stay of the
order of one as it seems to, this means that the {\it standard}
post-Newtonian approach, based on the standard Taylor approximants, is
probably very accurate. The post-Newtonian series is likely to
``converge well'', with a ``convergence radius'' of the order of
one\footnote{Actually, the post-Newtonian series could be only
asymptotic (hence divergent), but nevertheless it should give
excellent results provided that the series is truncated near some
optimal order of approximation. In this discussion we assume that the
3PN order is not too far from that optimum.}. Therefore the
order-of-magnitude estimates we proposed at the beginning of this
Section are probably correct. In particular the 3PN order should be
close to the ``exact'' solution even in the regime of the ICO.

Compare the situation with the case of Schwarzschild, for which we
have seen that $x^{\rm Sch}_{\rm ICO}=\frac{1}{6}$ is rather close to
the value of the convergence radius of the series given by $x^{\rm
Sch}_{\rm light-ring}=\frac{1}{3}$, hence the poor convergence of the
post-Newtonian expansion at the location of the ICO. By contrast, when
the two masses have the same size, we have $x_{\rm
ICO}(\frac{1}{4})\sim 0.2$ (see the Figure \ref{fig1}) which is quite
far from the ``convergence radius'' of the series, that we argued is
likely to be of the order of one. We thus expect the post-Newtonian
expansion (in the case of two black holes) to be well appropriate near
the ICO.

\section{Comparison with a numerical simulation}

We confront the prediction of the standard (Taylor-based)
post-Newtonian approach with a recent result of numerical relativity
by Gourgoulhon, Grandcl\'ement and Bonazzola\cite{GGB1,GGB2}. These
authors computed numerically the energy of binary black holes under
the assumptions of conformal flatness for the spatial metric and of
exactly circular orbits. The latter restriction is implemented by
requiring the existence of an ``helical'' Killing vector, time-like
inside the light cylinder associated with the circular motion and
space-like outside. In the numerical approach\cite{GGB1,GGB2} there
are no gravitational waves, the field is periodic in time, and the
gravitational potentials tend to zero at spatial infinity within a
restricted model equivalent to solving five out of the ten Einstein
field equations. Considering an evolutionary sequence of equilibrium
configurations Gourgoulhon {\it et al.}\cite{GGB1,GGB2} obtained
numerically the circular-orbit energy $E(\omega)$ and looked for the
innermost circular orbit or ICO.

The numerical calculation\cite{GGB1,GGB2} has been performed in the
case of {\it corotating} black holes, which are spinning with the
orbital angular velocity $\omega$. For the comparison we must
therefore include within our post-Newtonian formalism the effects of
spins appropriate to two Kerr black holes rotating at the orbital rate
$\omega$. The importance of the effect of spins in corotating systems
of neutron stars, for which the ICO is usually determined by the
hydrodynamical instability rather than by the effect of general
relativity, is well known\cite{DBSSU}. We expect that these effects
play some role in the case of black holes as well.

The total relativistic mass of the Kerr black hole is given
by\footnote{In the formulas (\ref{12})-(\ref{18}) we pose $G=1=c$.}

\begin{equation}\label{12}
m^2=m_{\rm irr}^2+\frac{S^2}{4m_{\rm irr}^2}\;,
\end{equation}
where $S$ is the spin, related to the usual Kerr parameter by $S=m a$,
and $m_{\rm irr}$ is the irreducible mass given by $m_{\rm
irr}=\frac{\sqrt{A}}{4\pi}$ ($A$ is the hole's surface area). The
angular velocity of the corotating black hole is $\omega =
\frac{\partial m}{\partial S}$ hence, from Eq. (\ref{12}),

\begin{equation}\label{13}
\omega = \frac{S}{2m^3\left[1+\sqrt{1-\frac{S^2}{m^4}}\right]}\;.
\end{equation}
Physically this angular velocity is the one of the outgoing photons
that remain for ever at the location of the light-like
horizon\cite{MTW}. Combining Eqs. (\ref{12}) and (\ref{13}) we obtain
$m$ and $S$ as functions of $m_{\rm irr}$ and $\omega$~:

\begin{eqnarray}\label{14}
m &=& \frac{m_{\rm irr}}{\sqrt{1-4m_{\rm irr}^2\,\omega^2}}\;,\\ S &=&
\frac{4m_{\rm irr}^3\omega}{\sqrt{1-4m_{\rm irr}^2\,\omega^2}}\;.
\end{eqnarray}
This is the right thing to do since $\omega$ is the basic variable
describing each equilibrium configuration calculated
numerically\cite{GGB1,GGB2}, and because the irreducible masses are
the ones which are held constant along the numerical evolutionary
sequences\cite{GGB1,GGB2}. In the limit of small rotations we have

\begin{equation}\label{15}
S = I\,\omega + {\cal O}\left(\omega^3\right)\;,
\end{equation}
where $I=4 m_{\rm irr}^3$ is the moment of inertia of the Kerr black
hole\cite{paradigm}. Next the total mass-energy is

\begin{equation}\label{16}
m = m_{\rm irr}+\frac{1}{2} I\,\omega^2 + {\cal
O}\left(\omega^4\right)\;.
\end{equation}
It involves the standard kinetic energy of the spin.

To take into account all the spin effects our first task is to replace
all the masses entering the energy function (\ref{2})-(\ref{3c}) by
their equivalent expressions in terms of $\omega$ and the two
irreducible masses. It is clear that the leading contribution is that
of the spin kinetic energy given by Eq. (\ref{16}) and comes from the
replacement of the rest mass-energy $M c^2$ (where $M=m_1+m_2$). From
Eq. (\ref{16}) this effect in the case of corotating binaries is of
order $\omega^2$, which means by comparison with
Eqs. (\ref{1})-(\ref{2}) that it is equivalent to an ``orbital''
effect at the 2PN order. Higher-order corrections in Eq. (\ref{16}),
which behave at least like $\omega^4$, will correspond at least to the
orbital 5PN order and are negligible for the present purpose. In
addition there will be a subdominant contribution, of the order of
$\omega^{8/3}$ equivalent to 3PN order, which comes from the
replacement of the masses into the ``Newtonian'' part, proportional to
$x\propto \omega^{2/3}$, of the energy $E$ [see Eq. (\ref{2})]. With
the 3PN accuracy we do not need to replace the masses that enter into
the post-Newtonian corrections in $E$, so these masses can be
considered to be the irreducible ones in these terms.

Our second task is to include the specific relativistic effects due to
the spins, namely the spin-orbit (S.O.) interaction and the spin-spin
(S.S.) one. In the case of spins $S_1$ and $S_2$ aligned parallel to
the orbital angular momentum (and right-handed with respect to the
sense of motion) the S.O. energy reads

\begin{equation}\label{17}
E_{\rm S.O.}= -\mu\, (M\omega)^{5/3}
 \Bigg[\left(\frac{4}{3}\frac{m_1^2}{M^2}+\nu\right)\frac{S_1}{m_1^2}+
 \left(\frac{4}{3}\frac{m_2^2}{M^2}+\nu\right)\frac{S_2}{m_2^2}\Bigg]\;.
\end{equation}
Here we are employing the formula given by Kidder, Will and
Wiseman\cite{KWWspin} who have computed the S.O. contribution as
expressed by means of the orbital frequency $\omega$ (this is what we
need in view of the comparison with the numerical
work\cite{GGB1,GGB2}). The basis of their computation is the work of
Barker and O'Connell\cite{BOC} who obtained the formula given in terms
of the orbital separation $r$. The derivation of Eq. (\ref{17}) in
Ref.\cite{KWWspin} takes correctly into account the fact that the
relation between the orbital separation $r$ (in a given coordinate
system) and the frequency $\omega$ depends on the spins. We
immediately infer from Eq. (\ref{15}) that in the case of corotating
black-holes the S.O. effect is equivalent to a 3PN orbital effect and
thus must be retained with the present accuracy [with this
approximation, the masses in Eq. (\ref{17}) are the irreducible
ones]. As for the S.S. interaction (still in the case of spins aligned
with the orbital angular momentum) it is given by

\begin{equation}\label{18}
E_{\rm S.S.} = \mu\,\nu\, (M\omega)^2 \frac{S_1\,S_2}{m_1^2\,m_2^2}\;.
\end{equation}
The S.S. effect can be neglected here because it is of the orbital
order 5PN for corotating systems.

Summaryzing, the contributions due to the corotating spins at the 3PN
order are three in all~: the main one is that of the spin kinetic
energy and arises at the 2PN order; then we have two subdominant
contributions at the 3PN order coming respectively from a mixing
between the spin kinetic energy and the Newtonian orbital energy, and
from the S.O. interaction (\ref{17}). We can neglect all the other
terms. Summing up these three contributions we find that the energy of
the corotating spins is (coming back to the notation of Section 2)

\begin{equation}\label{19}
\Delta E^{\rm corot}(x) = M \,\!c^2\,\!x \left\{
(2-6\nu)x^2+\left(-\frac{18}{3}\nu+13\nu^2\right)x^3+{\cal
O}(x^4)\right\}\;.
\end{equation}
The complete 3PN energy of the corotating binary is the sum of
Eqs. (\ref{2})-(\ref{3c}) and (\ref{19}). Notice that we must now
understand all the masses in (\ref{2})-(\ref{3c}) and (\ref{19}) as
being the irreducible masses --- we no longer indicate the
superscripts ``irr'' ---, which for the comparison with the numerical
work must be assumed to stay constant when the binary evolves.

\begin{figure}[htbp]
\centerline{\epsfxsize=10cm \epsfbox{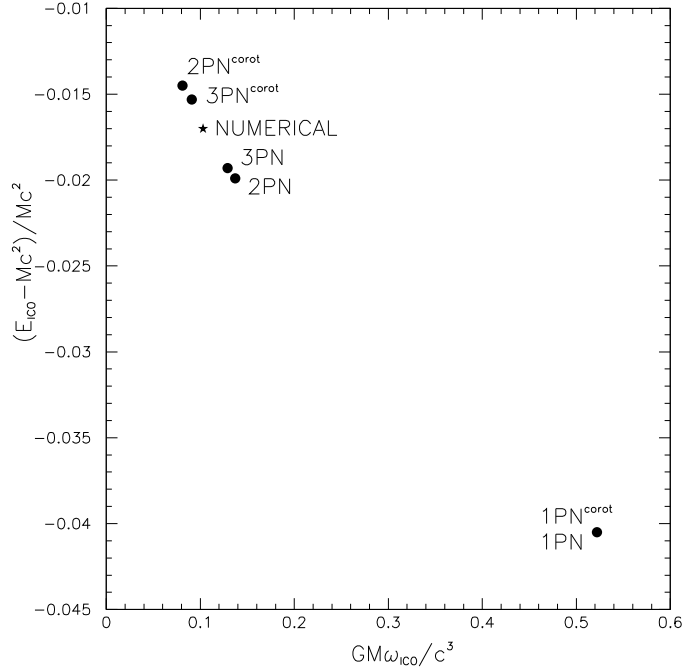}}
\vspace{0.5cm}
\caption{Results for the binging energy $E_{\rm ICO}-M c^2$ versus
$\omega_{\rm ICO}$ in the equal-mass case. The asterisk marks the
result calculated by numerical relativity. The points indicated by
1PN, 2PN and 3PN are computed from Eqs. (\ref{2})-(\ref{3c}), and
correspond to irrotational binaries. The points denoted by 1PN$^{\rm
corot}$, 2PN$^{\rm corot}$ and 3PN$^{\rm corot}$ come from the sum of
Eqs. (\ref{2})-(\ref{3c}) and (\ref{19}), and describe corotational
binaries. Both 3PN are 3PN$^{\rm corot}$ are shown for $\omega_{\rm
s}=0$.}
\label{fig1}
\end{figure}

The Figure \ref{fig1} (issued from the work\cite{B02ico}) presents
our results for $E_{\rm ICO}$ in the case of irrotational and
corotational binaries. Since $\Delta E^{\rm corot}$, given by
Eq. (\ref{19}), is at least of order 2PN, the result for 1PN$^{\rm
corot}$ is the same as for 1PN in the irrotational case; then,
obviously, 2PN$^{\rm corot}$ takes into account only the leading 2PN
corotation effect [i.e. the spin kinetic energy given by
Eq. (\ref{16})], while 3PN$^{\rm corot}$ involves also, in particular,
the corotational S.O. coupling at the 3PN order. In addition we
present in Figure \ref{fig1} the numerical point obtained by numerical
relativity under the assumptions of conformal flatness and of helical
symmetry\cite{GGB1,GGB2}. As we can see the 3PN points, and even the
2PN ones, are rather close to the numerical value. The fact that the
2PN and 3PN values are so close to each other is excellent, and
confirms the good accuracy of the approximation we concluded in
Section 3. In fact one might say that the role of the 3PN
approximation is merely to ``prove'' the value already given by the
2PN one (but of course, had we not computed the 3PN term, we would not
be able to trust very much the 2PN value). As expected, the best
agreement we obtain is for the 3PN approximation and in the case of
corotation~: i.e. the point 3PN$^{\rm corot}$. However, the 1PN
approximation is clearly not precise enough, but this is not
surprising in the highly relativistic regime of the ICO.

In conclusion, we find that the location of the ICO computed by
numerical relativity, under the helical-symmetry and
conformal-flatness approximations, is in good agreement with the
post-Newtonian prediction. (See Ref.\cite{DGG} for the results
calculated within the EOB method at the 3PN order, which are close to
the ones reported in Figure \ref{fig1}.) This constitutes an
appreciable improvement of the previous situation, because we recall
that the earlier estimates of the ICO in post-Newtonian
theory\cite{KWW} and numerical relativity\cite{Pfeiffer,Baumgarte}
strongly disagreed with each other, and do not match with the present
3PN results (see Ref.\cite{GGB2} for further discussion).

Notice in Figure \ref{fig1} the difference between the energies of the
corotational and irrotational configurations which amounts to
approximately $3.5\times 10^{-3}$ units of $M c^2$. We find that this
amount is mainly made of the (positive) kinetic energy of the spins,
which is about $+5.0\times 10^{-3}$, but that the S.O interaction at
the 3PN order, equal to $-1.3\times 10^{-3}$, reduces slightly the
effect, while the coupling between spin kinetic and orbital terms is
quite negligible, of the order of $-0.2\times 10^{-3}$. We feel that
these numerical values are physically satisfying, and well in
accordance with our order-of-magnitude estimates in Section 3 which
showed that in the regime of the ICO the post-Newtonian series
``converges well'', with the successive post-Newtonian approximations
becoming numerically smaller and smaller. We have checked that the
higher-order spin effects like the S.S. interaction [see
Eq. (\ref{18})] which arises at the 5PN order for corotating systems
are completely negligible numerically (the S.S. effect is about
$+0.02\times 10^{-3}$).

Recently Damour {\it et al.}\cite{DGG} computed the effects of spin in
corotating black hole binaries within the effective-one-body (EOB)
approach. They find a result which differs from our
computation\cite{B02ico} presented above. Basically they obtain that
the kinetic energy of the spins is about $+5.0\times 10^{-3}$, in
agreement with ours, but that the S.O. effect is much larger than in
our computation, and nearly compensates in their approach the spin
kinetic energy. The net result given in Ref.\cite{DGG} for the total
effect of spins is then about $+1\times 10^{-3}$, much smaller than
our own result. The surprising fact (in the author's opinion) is that
in the EOB method a relativistic effect like the S.O. coupling, which
according to the {\it a priori} expectation based on the estimates of
Section 3 should be small, can nearly cancel out a ``Newtonian'' term,
which one would physically consider to be the dominant one. Although
Refs.\cite{B02ico} and\cite{DGG} are in very good agreement for the
ICO of corotational binaries, and in rather good agreement for
irrotational binaries, their discrepancy concerning the numerical
value of the S.O. effect remains unexplained. But let us try a
guess. As we argued in Section 3 the resummation techniques are
probably justified only when the post-Newtonian coefficients in the
energy function get bigger and bigger as the order of approximation
increases. Our guess would be that perhaps the EOB method is somehow
forced to attribute in front of the S.O. term a big post-Newtonian
coefficient in order to maintain its internal consistency. With an
abnormally big coefficient one might explain the discrepancy.

\section*{Acknowledgements}
The author is grateful to an anonymous referee of his
paper\cite{B02ico} for pointing out the interest of looking at the
numerical values of the post-Newtonian coefficients.

%\pagebreak

%\narrowtext 

\end{document}